\documentclass[showpacs,twocolumn,pra]{revtex4}
\usepackage{amsfonts}
\usepackage{amsmath}
\usepackage{amssymb}
\usepackage{graphicx}

\begin{document}

\title{Spin dynamics and domain formation of a spinor Bose-Einstein
condensate in an optical cavity}

\begin{abstract}
We consider a ferromagnetic spin-1 Bose-Einstein condensate (BEC)
dispersively coupled to a unidirectional ring cavity. We show that the
ability of a cavity to modify, in a highly nonlinear fashion, matter-wave
phase shifts adds a new dimension to the study of spinor condensates both
within and beyond the single-mode approximation. In addition to
demonstrating strong matter-wave bistability as in our earlier publication
[L. Zhou \textit{et al}., Phys. Rev. Lett. \textbf{103}, 160403 (2009)], we
show that the interplay between atomic and cavity fields can greatly enrich
both the physics of critical slowing down in spin mixing dynamics and the
physics of spin-domain formation in spinor condensates.
\end{abstract}

\author{Lu Zhou$^{1}$, Han Pu$^{2}$, Hong Y. Ling$^{3}$, Keye Zhang$^{1}$
and Weiping Zhang$^{1}$}
\affiliation{$^{1}$State Key Laboratory of Precision Spectroscopy, Department of Physics,
East China Normal University, Shanghai 200062, China}
\affiliation{$^{2}$Department of Physics and Astronomy, and Rice Quantum Institute, Rice
University, Houston, TX 77251-1892, USA }
\affiliation{$^{3}$Department of Physics and Astronomy, Rowan University, Glassboro, New
Jersey 08028-1700, USA}
\pacs{03.75.Mn, 03.75.Kk, 42.50.Pq, 42.65.-k}
\maketitle

\section{introduction}

Experimental realization of spinor Bose-Einstein condensates (BEC) has
opened up a new research direction of cold atom physics \cite{ketterle98},
in which superfluidity and magnetism are simultaneously realized. Compared
to scalar condensates, spinor condensates possess unique features: (i) The
spin-dependent collision interactions allow for the population exchange
among hyperfine spin states; (ii) The spinor condensate is described by an
order parameter with vector character and therefore may exhibit spontaneous
magnetic ordering. These give rise to spin-dependent phenomena such as
coherent spin mixing, spin textures and vortices, spin waves and spin
domains. These phenomena have been extensively studied in theory \cite%
{pu00,you05,pu982,ho98,kivshar08,kivshar01,passos04,gu07,jordi09,damski07,santos07,ueda05,zhang05}
and demonstrated by a few pioneering experimental works \cite%
{chapman04,chapman05,lett07,ketterle03,stamper06,ketterle98,lett09,ketterle99}%
.

In the study of spinor BEC, it has been found that magnetic field plays an
important role, particularly via the quadratic Zeeman effect. Coherent
control of the spin-dependent behavior has been achieved by tuning magnetic
field. These include the control of the oscillation period and amplitude of
coherent spin mixing \cite%
{pu00,chapman04,chapman05,lett07,passos04,jordi09,you05}, formation of spin
domain structure \cite%
{ketterle98,stamper06,kivshar08,chapman05,kivshar01,gu07,ueda05,zhang05} and
quantum phase transitions between different magnetically ordered states \cite%
{stamper06,damski07,lett09}.

In another frontier of cold atom research, recent experimental progress have
realized strong coupling of BEC to electromagnetic modes of optical cavity
\cite{esslinger07,reichel07}. This heralds a new regime of cavity quantum
electrodynamics, where a cavity field at the level of a single photon can
significantly affect the collective motion of the atomic samples, hence
opening up new possibilities in manipulating ultracold atomic gases with
cavity-mediated nonlinear interaction. Previous works focused on the
interplay between the cavity field and the atomic external degrees of
freedom --- the center-of-mass motion of scalar condensates \cite%
{esslinger08,kurn07,kurn08,ritsch00,zhang08,domokos02,larson08,meystre10,moore99}%
. The ground state and collective excitations \cite{ritsch00,domokos02},
cavity induced Mott insulator-superfluid phase transition \cite{larson08}
and cavity optomechanics \cite{meystre10} were theoretically investigated in
detail. Such a system was also shown to have the potential applications in
probing atomic quantum statistics in optical lattices and atomic quantum
state preparation \cite{mekhov07}. Experimentally, optical bistability at
few-photon level has been observed, which is made possible by the strong
atom-photon coupling \cite{kurn07,esslinger08}.

In our recent work \cite{zhou09}, a system of a spin-1 BEC trapped inside a
unidirectional ring cavity was studied, where the cavity couples to the
atomic internal spin degrees of freedom. We examined the equilibrium
properties of this system under the single-mode approximation (SMA) and
showed that the interplay between the atomic spin mixing and the cavity
light field can lead to strong matter-wave and optical bistability
simultaneously. Our current work is an extension of Ref.~\cite{zhou09}. Here
we will conduct a more complete investigation by including the study on the
non-equilibrium properties and the collective excitations of the system. We
will also examine the validity of the SMA and show that, when SMA becomes
invalid, spatial domain structure will form in the spinor condensate. This
study will help us gain insight into such properties as the spinor dynamics,
dynamical stability, spin domain formation, etc.

The rest of the paper is organized as follows. Section II introduces the
theoretical model. Section III is devoted to a discussion of spinor dynamics
under the SMA, where both equilibrium and non-equilibrium properties are
studied. The validity of the SMA is examined in Sec. IV by investigating the
modulational stability of a homogeneous system. We then present results
showing the formation of spin domain structure in the ground state in the
regime where the SMA becomes invalid. Finally we conclude in Sec. V.

\section{model}

\begin{figure}[tbh]
\includegraphics[width=8cm]{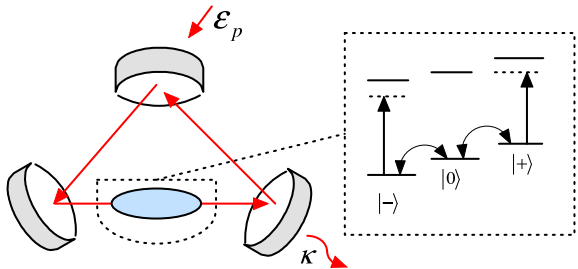}
\caption{{\protect\footnotesize (color online) Schematic diagram showing the
system under consideration. An }$F=1${\protect\footnotesize \ spinor
condensate is trapped inside the cavity using an optical dipole trap. The
population of different spin components can exchange via spin mixing. The
cavity is coherently driven by an external laser with amplitude }$\protect%
\varepsilon _{p}${\protect\footnotesize \ and decays with a rate }$\protect%
\kappa ${\protect\footnotesize . The cavity field is }$\protect\pi $%
{\protect\footnotesize \ polarized and is dispersively coupled to the atomic
system.}}
\label{schematic diagram}
\end{figure}

Figure \ref{schematic diagram} is a schematic of our model in which a spinor
BEC with hyperfine spin $F_{g}=1$ confined in an optical dipole trap by a
spin-independent trapping potential $V_{T}\left( \mathbf{r}\right) $ is
placed inside a unidirectional ring cavity. The cavity is driven by a
coherent laser field with amplitude $\varepsilon _{p}$ and frequency $\omega
_{p}$, far detuned away from the $F_{g}=1\leftrightarrow F_{e}=1$ transition
frequency $\omega _{a}$ such that the atom-photon interaction is essentially
of dispersive nature. Further, the cavity is assumed to support a single $%
\pi $-polarized electromagnetic mode characterized by a frequency $\omega
_{c}$, a decay rate $\kappa $. The selection rule then allows states $%
\left\vert F_{g}=1,m_{g}=\pm 1\right\rangle $ to be coupled to the
corresponding states in the excited manifold with the same magnetic quantum
numbers $\left\vert F_{e}=1,m_{e}=\pm 1\right\rangle $, while forbids state $%
\left\vert F_{g}=1,m_{g}=0\right\rangle $ to make dipole transitions to any
excited states as shown in Fig. \ref{schematic diagram}. However, population
in state $\left\vert F_{g}=1,m_{g}=0\right\rangle $ can be distributed to
other sublevels via the two-body $s$-wave spin exchange collisions. The
total $s$-wave interaction are described by two numbers, $c_{0}=4\pi \hbar
^{2}\left( 2a_{2}+a_{0}\right) /3m_{a}$ and $c_{2}=4\pi \hbar ^{2}\left(
a_{2}-a_{0}\right) /3m_{a}$, where $m_{a}$ is the atom mass, and $a_{f}$ is
the s-wave scattering lengths in the hyperfine channel with a total spin $%
f=0 $ or $2$. Since the antiferromagnetic case of $^{23}$Na has been
considered in our previous work \cite{zhou09}, here we will focus on the
ferromagnetic case of $^{87}$Rb spin-1 BEC where $c_2<0$.

In this work, we take the standard mean-field approach, describing the
cavity field with a complex amplitude $\alpha \left( t\right) $ which
amounts to assuming the cavity field to be represented by a coherent state,
and the spinor condensate with the order parameters $\psi _{-}\left( \mathbf{%
r},t\right) ,\psi _{0}\left( \mathbf{r},t\right) ,$ and $\psi _{+}\left(
\mathbf{r},t\right) $, which represent the wavefunctions in magnetic
sublevels $m_{g}=-1,0,$ and $+1$, respectively. This treatment is justified
when the condensate atom number $N_{\alpha }=\int n_{\alpha }\left( \mathbf{r%
}\right) d\mathbf{r}$ (where $n_{\alpha }=\left\vert \psi _{\alpha
}\right\vert ^{2}$ is the atom number density) in magnetic sublevel $\alpha $
are sufficiently large. The equations of motion then read
\begin{subequations}
\label{general equation}
\begin{align}
i\hbar \dot{\psi}_{\pm }& =\left[ \mathcal{L}+U_{0}\left\vert \alpha
\right\vert ^{2}+c_{2}\left( n_{\pm }+n_{0}-n_{\mp }\right) \right] \psi
_{\pm }+c_{2}\psi _{0}^{2}\psi _{\mp }^{\ast },  \label{gea} \\
i\hbar \dot{\psi}_{0}& =\left[ \mathcal{L}+c_{2}\left( n_{+}+n_{-}\right) %
\right] \psi _{0}+2c_{2}\psi _{+}\psi _{-}\psi _{0}^{\ast },  \label{geb} \\
\dot{\alpha}& =\left[ i\delta _{c}-iU_{0}\left( N_{+}+N_{-}\right) -\kappa %
\right] \alpha +\varepsilon _{p},  \label{gec}
\end{align}%
where $\mathcal{L}=\hat{p}^{2}/2m_{a}+V_{T}\left( \mathbf{r}\right) +c_{0}n$
is the spin-independent part of the Hamiltonian, $n=n_{+}+n_{0}+n_{-}$ is
the total atomic density, $\ \delta _{c}=\omega _{p}-\omega _{c}$ is the
cavity detuning relative to the external laser field, and $%
U_{0}=g^{2}/\left( \omega _{p}-\omega _{a}\right) $ is the effective
atom-photon coupling, with $g$ being atom-cavity mode coupling constant. \
Further, since the cavity decay rate $\kappa $ is typically much larger than
the spin oscillation frequency, in what follows, we adiabatically eliminate $%
\alpha $ from Eq. (\ref{gec}), replacing $\alpha $ in Eq. (\ref{gea}) with
\end{subequations}
\begin{equation}
\alpha \left( t\right) =\frac{\varepsilon _{p}}{\kappa -i\left[ \delta
_{c}-U_{0}\left( N_{+}+N_{-}\right) \right] } \,.  \label{cavity}
\end{equation}

One may immediately observe from Eqs.~(\ref{general equation}) that the
dispersive interaction between cavity photons and the condensate atoms
introduces an effective quadratic Zeeman energy shift, $U_{0}|\alpha |^{2}$,
to $m_{g}=\pm 1$ states relative to the $m_{g}=0$ state. However, unlike the
Zeeman shift due to an external magnetic field or to a strong off-resonant
laser field \cite{santos07}, a key feature of this effective shift is that
it is sensitive to the spin population distribution of the condensate, as
manifested by Eq.~(\ref{cavity}). As such, it generates a new effective
spin-dependent interaction which in turn induces a new set of nonlinear
phenomena in spinor condensate. In what follows, we will describe in detail
such new phenomena.


\section{spin dynamics under SMA}

In this section, we consider the spin dynamics under the assumption of SMA.
This describes, for example, a condensate whose size is smaller than the
spin healing length $\xi _{s}$ defined as $\xi _{s}=h/\sqrt{2m_{a}\left\vert
c_{2}\right\vert n}$ which represents a length scale over which a local
perturbation in spin density gets forgotten. Under the SMA, each spin
component shares the same spatial wavefuntion $\phi \left( \mathbf{r}\right)
$ according to%
\begin{equation}
\psi _{\alpha }\left( \mathbf{r},t\right) =\sqrt{N}\phi \left( \mathbf{r}%
\right) \sqrt{\rho _{\alpha }}\exp \left[ -i\left( \mu t+\theta _{\alpha
}\right) \right] ,\text{ }\alpha =\pm ,0,  \label{sma}
\end{equation}%
where $\theta _{\alpha }$ is the phase, $\rho _{\alpha }$ is the population
normalized with respect to the total atom number $N=\sum_{\alpha }N_{\alpha
} $, and\ $\phi \left( \mathbf{r}\right) $ is the solution to the
time-independent Gross-Pitaevskii equation: $\mathcal{L}\phi =\mu \phi $,
where $\mu $ is the chemical potential\ and $\phi \left( \mathbf{r}\right) $
satisfy the normalization condition $\int d\mathbf{r}\left\vert \phi \left(
\mathbf{r}\right) \right\vert ^{2}=1$.

By inserting Eq. (\ref{sma}) into Eqs. (\ref{gea}) and (\ref{geb}), we
arrive at a set of equations
\begin{subequations}
\label{classical equation}
\begin{align}
\frac{d\rho _{0}}{d\tau }& =2\lambda _{a}\rho _{0}\sqrt{\left( 1-\rho
_{0}\right) ^{2}-m^{2}}\sin \theta ,  \label{cea} \\
\frac{d\theta }{d\tau }& =-2\frac{\bar{U}_{0}\left\vert \alpha \right\vert
^{2}}{N}+2\lambda _{a}\times  \notag \\
& \left[ 1-2\rho _{0}+\frac{\left( 1-\rho _{0}\right) \left( 1-2\rho
_{0}\right) -m^{2}}{\sqrt{\left( 1-\rho _{0}\right) ^{2}-m^{2}}}\cos \theta %
\right] ,  \label{ceb}
\end{align}%
which describe the dynamics of a mixed state in which none of the spin
component vanishes, where $\theta =2\theta _{0}-\theta _{+}-\theta _{-}$ is
the relative phase, $m=\rho _{+}-\rho _{-}$ the magnetization, and $\tau
=\kappa t$ the dimensionless time. In Eqs. (\ref{classical equation}), we
have also introduced other dimensionless quantities given by
\end{subequations}
\begin{equation*}
\lambda _{a}=\frac{Nc_{2}\int d\mathbf{r}\left\vert \phi \left( \mathbf{r}%
\right) \right\vert ^{4}}{\kappa },\text{ }\bar{U}_{0}=\frac{NU_{0}}{\kappa }%
,\text{ }\eta =\frac{\varepsilon _{p}}{\kappa },\text{ }\bar{\delta}_{c}=%
\frac{\delta _{c}}{\kappa }.
\end{equation*}%
To facilitate our study below, we follow Refs. \cite{you05,zhou07} and use $%
d\rho _{0}/d\tau =-2\partial H/\partial \theta $ and $d\theta /d\tau
=2\partial H/\partial \rho _{0}$ to construct, in terms of two conjugate
variables $\rho _{0}$ and $\theta $, the following mean-field Hamiltonian $H$%
\begin{equation}
H=\lambda _{a}\rho _{0}\left[ 1-\rho _{0}+\sqrt{\left( 1-\rho _{0}\right)
^{2}-m^{2}}\cos \theta \right] +U\left( \rho _{0}\right) ,
\label{hamiltonian}
\end{equation}%
where
\begin{equation*}
U\left( \rho _{0}\right) =\frac{\eta ^{2}}{N}\arctan \left[ \bar{U}%
_{0}\left( 1-\rho _{0}\right) -\bar{\delta}_{c}\right]
\end{equation*}%
represents the cavity-mediated atom-atom interaction.

\subsection{Equilibrium Property: Bistability}

In this subsection, we will use Eqs. (\ref{classical equation}) to study the
equilibrium property of a condensate in the parameter regime that supports
bistability. As can be seen from Eq.~(\ref{cea}), at steady state, there are
two branches of stationary solutions: one with $\theta =0$ (the in-phase
state) and the other with $\theta =\pi $ (the out-of-phase state). The
in-phase state always has a lower energy for $c_{2}<0$ and we will therefore
only focus on the in-phase state in this work. In addition, we will restrict
ourselves to the case with zero magnetization $m=0$, i.e., we only consider
the case where there are equal number of $m_g=1$ and $m_g=-1$ atoms.

Under these conditions, the intracavity photon number can be found, by
combing the stationary solution of Eq.~(\ref{ceb}) with Eq.~(\ref{cavity}),
to obey the following transcendental equation
\begin{equation*}
\left\vert \alpha \right\vert ^{2}=\frac{\eta ^{2}}{1+\left( \Delta +\chi
\left\vert \alpha \right\vert ^{2}\right) ^{2}},
\end{equation*}%
where $\Delta =\bar{U}_{0}/2-\bar{\delta}_{c}$ and $\chi =\bar{U}%
_{0}^{2}/4N\lambda _{a}$. It is well-known that when $\eta ^{2}\left\vert
\chi \right\vert >8\sqrt{3}/9$, the system will display bistable behavior
\cite{boyd}.

\begin{figure}[tbh]
\includegraphics[width=7cm]{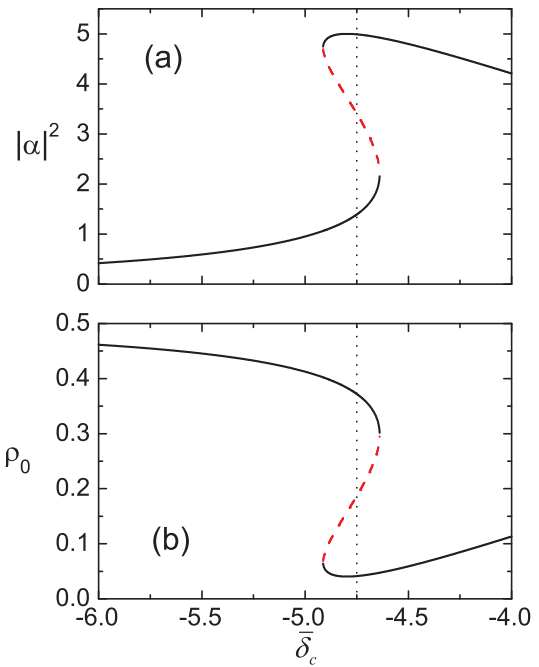}
\caption{{\protect\footnotesize (color online) (a) Mean intracavity photon
number }$\left\vert \protect\alpha \right\vert ^{2}${\protect\footnotesize \
and (b) the normalized spin-0 population }$\protect\rho _{0}$%
{\protect\footnotesize \ versus cavity-pump detuning }$\bar{\protect\delta}%
_{c}${\protect\footnotesize \ for a steady-state solution with }$\protect%
\theta =0${\protect\footnotesize . The ones represented by the red dashed
lines correspond to dynamically unstable solutions. The vertical dotted
lines indicate the position of the first-order transition which occurs at }$%
\bar{\protect\delta}_{c}=-4.75${\protect\footnotesize .}}
\label{bistability}
\end{figure}
Figure \ref{bistability}(a) shows how the intracavity photon number changes
with detuning $\bar{\delta}_{c}$, based on a set of realistic parameters: $%
\lambda _{a}=-6.8\times 10^{-5}$ \cite{lambda}, $\bar{U}_{0}=-5$, $\eta
^{2}=5$, and $N=2\times 10^{5}$. With this set of parameters, $\eta
^{2}\left\vert \chi \right\vert $ is found to be around $2.3$, which is
above the threshold value $8\sqrt{3}/9\approx 1.54.$ Indeed, for $-4.9<\bar{%
\delta}_{c}<-4.6$, the system supports three stationary solutions. The
dynamical properties of these solutions can be studied with the standard
linear stability analysis. Substituting $\rho _{0}=\rho _{0}^{s}+\delta \rho
_{0}$ and $\theta =\theta ^{s}+\delta \theta $ ($\rho _{0}^{s}$ is the
stationary solution with $\theta ^{s}=0$) into Eqs.~(\ref{classical equation}%
) and keeping terms up to the first order in fluctuations $\left( \delta
\rho _{0},\delta \theta \right) $, we have%
\begin{align*}
\frac{d}{d\tau }\delta \rho _{0}& =2\lambda _{a}\rho _{0}^{s}\left( 1-\rho
_{0}^{s}\right) \delta \theta , \\
\frac{d}{d\tau }\delta \theta & =-2\left( 4\lambda _{a}+\frac{\bar{U}_{0}}{N}%
\left. \frac{\partial \left\vert \alpha \right\vert ^{2}}{\partial \rho _{0}}%
\right\vert _{\rho _{0}=\rho _{0}^{s}}\right) \delta \rho _{0},
\end{align*}%
from which we find the small oscillation frequency $\omega $ as determined
by the following equation%
\begin{equation*}
\omega ^{2}=4\lambda _{a}\rho _{0}^{s}\left( 1-\rho _{0}^{s}\right) \left(
4\lambda _{a}+\left. \frac{\bar{U}_{0}}{N}\frac{\partial \left\vert \alpha
\right\vert ^{2}}{\partial \rho _{0}}\right\vert _{\rho _{0}=\rho
_{0}^{s}}\right) .
\end{equation*}%
In order to assure the dynamical stability of the system, $\omega ^{2}$
should be positive. We find that in the region with three solutions, two of
them are dynamically stable while the third one is dynamically unstable.
This unstable state is shown by the the red dashed line in Fig.~\ref%
{bistability}, it links the two stable ones, representing a typical example
of bistability.

In the region where the intracavity photon number is low, the interaction is
dominated by the intrinsic $s$-wave scattering, which favors the
ferromagnetic state in which $\rho _{0}=0.5$ for $m=0$. In the region where
the photon number is high, the cavity-induced effective Zeeman effect takes
a more prominent role which, for the choice of $U_{0}<0$, favors a
condensate in the $m_{g}=\pm 1$ magnetic sublevels in which $\rho _{0}$
becomes small. If $\alpha $ is fixed to a value independent of the atomic
dynamics as in the case when it represents a strong off-resonant laser field
\cite{santos07}, the system will experience a smooth crossover from the
ferromagnetic interaction dominated phase to the Zeeman effect dominated
phase as the strength of $U_{0}$ is tuned. In our case, however, there is a
first-order transition located within the bistable region as indicated in
Fig.~\ref{bistability}. This phase transition exists as a result of the
cavity-mediated nonlinear atom-atom interaction.

\subsection{Non-equilibrium property: Critical Slowing Down}

In this subsection, we study the spin-mixing dynamics of the system
initially prepared in a state away from equilibrium. To begin with, we make
use of Eq.~(\ref{hamiltonian}) and rewrite Eq.~(\ref{cea}) for $m=0$ as%
\begin{equation}
\left( \frac{d\rho _{0}}{d\tau }\right) ^{2}=8\lambda _{a}\rho _{0}\left(
1-\rho _{0}\right) \left[ H-U\left( \rho _{0}\right) \right] -4\left[
H-U\left( \rho _{0}\right) \right] ^{2},  \label{nonequilibrium}
\end{equation}%
where $H$ is the energy of the system which is a constant determined by the
initial condition. In the cavity-free model when $U$ represents a constant
quadratic Zeeman shift independent of $\rho_0$, Eq.~(\ref{nonequilibrium})
is known to support analytical solutions in the form of elliptic functions
\cite{you05}. In our case, we have to resort to numerics to solve the above
equation. As the system is conserved, the spin dynamics is expected to
feature periodic population exchanges among different spin states, as in the
cavity-free model with a homogeneous magnetic field \cite{you05,passos04}.

\begin{figure}[tbh]
\includegraphics[width=9cm]{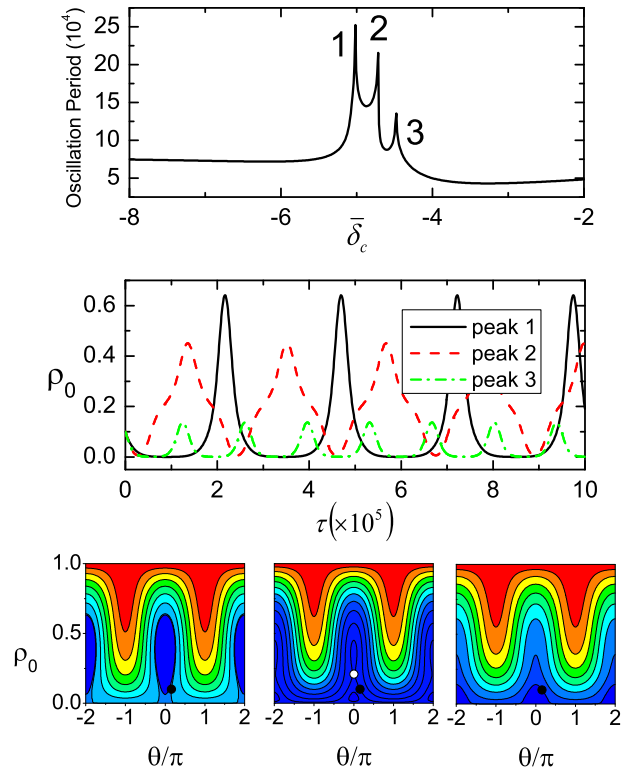}
\caption{{\protect\footnotesize (Color online) Upper panel: Period of spin
oscillations as a function of cavity-pump detuning }$\bar{\protect\delta}%
_{c} ${\protect\footnotesize . Middle panel: The anharmonic time evolution
of }$\protect\rho _{0}${\protect\footnotesize \ for the three peaks marked
in the upper panel. Lower panel: From left to right, the phase-space contour
plot of }$H${\protect\footnotesize \ corresponding, respectively, to the
peak 1, 2 and 3 marked in the upper panel. The black dots refer to the
initial state of the system, while the white dots refer to dynamically
unstable fixed points.}}
\label{oscillation}
\end{figure}

Figure \ref{oscillation} shows how the oscillation period changes with
cavity detuning $\bar{\delta}_{c}$, where the period is obtained by solving
Eq.~(\ref{nonequilibrium}) numerically starting from the initial condition $%
\left( \rho _{0}=0.1,\theta =0.16\pi \right) $ under the same set of
parameters that resulted in the equilibrium state in Fig. (\ref{bistability}%
) with $\theta =0$. Here, cavity detuning $\bar{\delta}_{c}$ serves as a
control knob with which the departure between the initial non-equilibrium
state $\left( \rho _{0}=0.1,\theta =0.16\pi \right) $ and the closest
equilibrium state (an in-phase state with $\theta =0$) can be conveniently
tuned. It plays a similar role as the magnetic field in the study of spin
dynamics in the presence of a homogeneous magnetic field. In the
ferromagnetic case, it has been theoretically predicted \cite{you05} that
there is a \textit{single} critical magnetic field around which oscillation
period diverges. In contrast, the period as a function of $\bar{\delta}_{c}$
in Fig.~\ref{oscillation} exhibits three peaks around which the period (or
the oscillation) experiences a dramatic enhancement (or slowing down)
\footnote[1]{%
This critical slowing down should not be confused with those discussed in
\cite{meystre79}, in which it refers to an extremely slow return of the
system to equilibrium in the vicinity of the bistable transition points.}.
The spin population $\rho _{0}$ as functions of time at three peaks are
illustrated in Fig.~\ref{oscillation}.

To gain physical insights into these dynamics, we plot in the bottom of Fig.~%
\ref{oscillation} the corresponding equal-$H$ contour diagrams in the phase
space defined by the conjugate pair $\left( \theta \text{, }\rho _{0}\right)
$. In a dissipationless system like ours, no matter how complicated the
system dynamics may look in the time domain, it evolves along one such
contour determined by the initial state (marked as a black dot in Fig.~\ref%
{oscillation}). The critical slowing down takes place when the energy
approaches a critical value $H_{c}$ below which the contour changes its
topology from an open to a closed line. In the pendulum analogy, it
corresponds to the pendulum approaching the vertical upright position. The
existence of a bistable region in our example makes the phenomenon of
critical slowing down far richer. As can be seen, both the first and third
peaks are located outside the bistable region, where only one attractor
representing the stable state at $\theta =0$ exists, while the second one is
inside the bistable region, where an unstable saddle point marked by a white
dot coexists with two attractors at $\theta =0$. Our results show that the
oscillation period strongly depends on the cavity light field, the pump
field can thus serve as a control knob for the spin-mixing dynamics.

\section{Beyond SMA}

So far we have focused our discussion within the SMA. In this section, we
will investigate the validity of the SMA and study the properties of the
system when the SMA becomes invalid.

\subsection{Modulational Instability of a Homogeneous Condensate}

In order to gain some physical insights into the validity of the SMA, we
first consider the case without the trapping potential and assume that the
condensate inside the cavity is homogeneous. In this case we have $\psi
_{\alpha }=\sqrt{n_{\alpha }}\exp \left( -i\mu _{\alpha }t-i\theta _{\alpha
}\right) $, where the atomic density $n_{\alpha }$ now becomes
position-independent, and the stationary solution $\left( n_{\alpha }^{s}%
\text{, }\theta _{\alpha }^{s}\right) $ is still determined by Eqs. (\ref%
{classical equation}) at steady state except that $\lambda _{a}$ should be
redefined as $\lambda _{a}\equiv c_{2}n/\kappa $.

In order to check whether these homogeneous states are stable against
spatial modulation, we examine the Bogoliubov collective excitation spectrum
by introducing small fluctuations around the steady-state solution.
Inserting $\psi _{\alpha }=\left( \sqrt{n_{\alpha }^{s}}+\delta \psi
_{\alpha }\right) \exp \left( -i\mu _{\alpha }^{s}t-i\theta _{\alpha
}^{s}\right) $ into Eqs.~(\ref{general equation}), where $\delta \psi
_{\alpha }$ can be expanded in momentum space as $\delta \psi _{\alpha
}\left( \mathbf{r},t\right) =\sum_{\mathbf{k}}\left[ u_{\alpha }\left(
t\right) \exp \left( i\mathbf{k}\cdot \mathbf{r}\right) +v_{\alpha }^{\ast
}\left( t\right) \exp \left( -i\mathbf{k}\cdot \mathbf{r}\right) \right] $,
we obtain a matrix equation $id\mathbf{x}/dt=\mathcal{M}\mathbf{x}$ for
vector $\mathbf{x}=\left( u_{+},u_{0},u_{-},v_{+}^{\ast },v_{0}^{\ast
},v_{-}^{\ast }\right) ^{T}$ where $\mathcal{M}$ is a matrix given in the
Appendix. The Bogoliubov modes are then given by the eigenvalue equations $%
\mathcal{M}\mathbf{x}=\hbar \omega \mathbf{x}$, where $\omega $ represents
the excitation frequency if it is real and signals modulational instability
with a growth rate $\mathrm{Im}\left( \omega \right) $ if it is complex.

\begin{figure}[tbh]
\includegraphics[width=8cm]{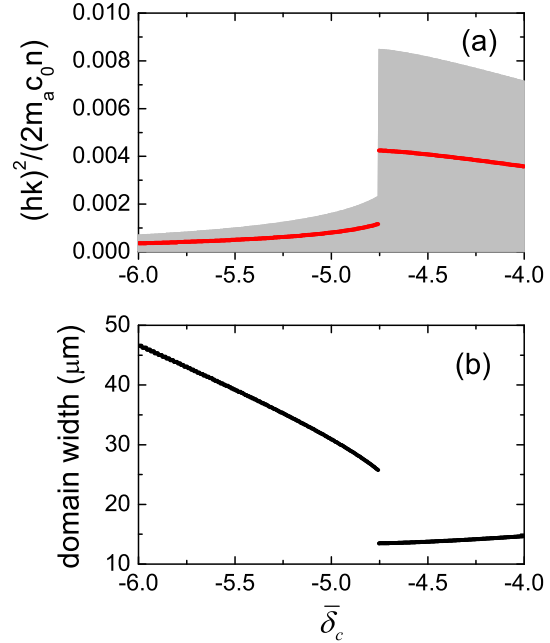}
\caption{{\protect\footnotesize (color online) (a) The grey area shows the
range of wavevector }$k${\protect\footnotesize \ corresponding to the
unstable excitations of the self-consistent ground-state of a homogeneous
rubidium condensate-cavity interacting system. The red lines refer to those
with the maximum instablity growth rate. (b) Spin domain width versus
cavity-pump detuning }$\bar{\protect\delta}_{c}${\protect\footnotesize .}}
\label{excitation}
\end{figure}

In the absence of cavity, the homogeneous ground state of a ferromagnetic $%
^{87}$Rb condensate is stable against spatial modulation even with a finite
magnetic field \cite{kivshar08}. This, as we shall show, will not be the
case when the cavity is introduced. To illustrate this, we numerically
diagonalize $\mathcal{M}$ to investigate the properties of Bogoliubov
excitations. For the spin-1 system as we considered here, there will be
three branches of Bogoliubov excitations --- two gapless branches and one
gapped branch. Our numerical calculations reveal that one of the gapless
branches will become unstable for certain values of the wavevector $k$.
Figure~\ref{excitation}(a) shows the range of unstable excitations
associated with $\left\vert k\right\vert \in \left[ 0,k_{\mathrm{m}}\right] $%
, and those with the maximum instability growth rate are represented by the
red lines. Furthermore, the Bogoliubov eigenvectors of these most unstable
modes are found to take the following form%
\begin{equation*}
u_{\alpha }^{T},v_{\alpha }^{T}\propto \left( -0.5,0,0.5\right) \text{ or }%
\left( 0.5,0,-0.5\right) ,
\end{equation*}%
which describe the spin waves with spin angular momentum $\pm \hbar $. The
exponential growth of these modes tends to induce spontaneous magnetization,
and spin domain will be formed as a result of the competition between local
spontaneous magnetization and the conservation of the total magnetization.
The size of the spin domain may be estimated by the inverse of the
wavenumber $2\pi /k_{\mathrm{m}}$, which is plotted in Fig.~\ref{excitation}%
(b).

It is important to note that if the total size of the condensate is small
compared to the domain width estimated above, the instability will be
suppressed.

\subsection{Spin Domain Structure}

Equipped with the insights gained from the study of a homogeneous condensate
in the previous subsection, we are now in the position to explore the effect
of cavity-induced atom-atom interaction on spin-domain formation in a
trapped condensate. For simplicity, we consider a cigar-shaped trap with a
harmonic trap potential $V_{T}\left( \mathbf{r}\right) =m_{a}\left[ \omega
_{\perp }^{2}\left( x^{2}+y^{2}\right) +\omega _{z}^{2}z^{2}\right] /2$ in
which the transverse trap frequency $\omega _{\perp }$ is much higher than
the longitudinal trap frequency $\omega _{z}$. This allows us to introduce a
longitudinal wavefunction $\phi _{\alpha }\left( z,t\right) $ via the ansatz
$\psi _{\alpha }\left( \mathbf{r},t\right) =\phi _{\perp }\left( x,y\right)
\phi _{\alpha }\left( z,t\right) \exp \left( -2i\omega _{\perp }t\right) $,
assuming that the transverse wavefunction $\phi _{\perp }\left( x,y\right) $
always remains in the ground state of the transverse potential. Following
the standard approach (see, for example, Ref.~\cite{kivshar01,kivshar08}),
we simplify Eqs.~(\ref{gea}), (\ref{geb}) and (\ref{cavity}) into a set of
equations for $\phi _{\alpha }\left( z,t\right) $
\begin{subequations}
\label{reduced equations}
\begin{align}
i\hbar \dot{\phi}_{\pm }& =\left[ \mathcal{\tilde{L}}+U_{0}\left\vert \alpha
\right\vert ^{2}+\bar{c}_{2}\left( \rho _{\pm }+\rho _{0}-\rho _{\mp
}\right) \right] \phi _{\pm }  \notag \\
& +\bar{c}_{2}\phi _{0}^{2}\phi _{\mp }^{\ast },  \label{rea} \\
i\hbar \dot{\phi}_{0}& =\left[ \mathcal{\tilde{L}}+\bar{c}_{2}\left( \rho
_{+}+\rho _{-}\right) \right] \phi _{0}+2\bar{c}_{2}\phi _{+}\phi _{-}\phi
_{0}^{\ast },  \label{reb}
\end{align}%
which describe an effective 1D trapped system, where
\end{subequations}
\begin{equation*}
\mathcal{\tilde{L}}=-\frac{\hbar ^{2}}{2m_{a}}\frac{\partial ^{2}}{\partial
z^{2}}+\frac{m}{2}\omega _{z}^{2}z^{2}+\bar{c}_{0}\rho ,
\end{equation*}%
with $\rho _{\alpha }=\left\vert \phi _{\alpha }\right\vert ^{2}$, $\rho
=\rho _{+}+\rho _{0}+\rho _{-}$, and $\bar{c}_{0\left( 2\right) }=c_{0\left(
2\right) }m_{a}\omega _{\perp }/2\pi \hbar $.

\begin{figure}[tbh]
\includegraphics[width=9cm]{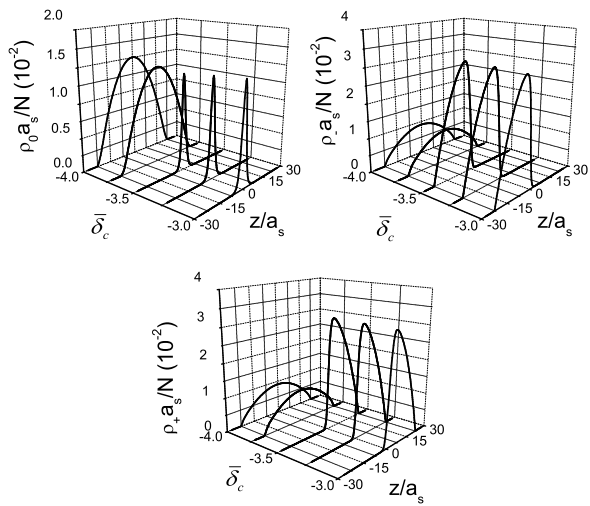}
\caption{{\protect\footnotesize The ground denstiy profile of a }$^{87}$%
{\protect\footnotesize Rb condensate trapped in a unidirectional ring
cavity. Here the distance }$z${\protect\footnotesize \ is scaled with }$%
a_{s}=\protect\sqrt{\hbar /m_{a}\protect\omega _{z}}${\protect\footnotesize %
, and the parameters used are specified in the main text.}}
\label{domain}
\end{figure}

In our calculation, we set the trap frequencies as $\omega _{\perp }=\left(
2\pi \right) 240$ Hz and $\omega _{z}=\left( 2\pi \right) 24$ Hz \cite%
{zhang05}, and other parameters same as before. The Thomas-Fermi radius in
the $z$-direction is then about $24$ $\mu m,$. In the numerical simulation,
we obtain the ground state in a self-consistent manner by propagating Eqs.~(%
\ref{reduced equations}) in imaginary time subject to the constraints set by
the conservation of both the total particle number and the magnetization.
The results are shown in Fig.~\ref{domain}. From the numerical simulation we
find there exists a critical value of the cavity-pump detuning $\Delta
_{c}\approx -3.5$. The ground state exhibits a typical spin domain structure
when $\bar{\delta}_{c}>\Delta _{c}$, in which $m_{g}=1$ and $-1$ states
occupy the opposite ends of the longtitudial trap. While for $\bar{\delta}%
_{c}<\Delta _{c}$, all three spin components are completely miscible with no
spin domains forming, and the ground state is well described by the SMA.
This is a clear proof that the cavity light field can be used to control
spin domain formation in the condensate. The mechanism lies in the fact that
the domain width can be significantly modified by tuning the cavity-pump
detuning $\bar{\delta}_{c}$, as we have shown in Fig.~\ref{excitation}(b).
When $\bar{\delta}_{c}>\Delta _{c}$, the domain width (around $14$ $\mu m$)
is smaller than the condensate size and spin domain can be formed. While for
$\bar{\delta}_{c}<\Delta _{c}$, the domain width is larger than the size of
the condensate, then the domain formation instability is suppressed.

At this point, we comment that spin domains was first observed in the
ground-state of a $^{23}$Na antiferromagnetic condensate in the presence of
the magnetic field gradient \cite{ketterle98}. Later studies \cite%
{kivshar01,zhang05,ueda05} discovered that a ferromagnetic spinor condensate
initially prepared in an excited state will be subject to dynamical
instability and lead to spin domain formation, while antiferromagnetic ones
are dynamically stable. The experiment of Ref.~\cite{stamper06} displayed
the spin domains formation in a quenched $^{87}$Rb ferromagnetic condensate.
Recent work \cite{kivshar08} clarified that for a spin-1 condensate subject
to a homogeneous magnetic field, the ground state exhibits domain formation
only in antiferromagnetic condensates, but not in the ferromagnetic ones.
The significance of our work here is that spin domain structures can also be
created in the ground state of a ferromagnetic condensate with the aid of a
cavity. This can be traced to the effective spin-dependent atom-atom
interation induced by the cavity.

\section{conclusion}

In conclusion, we have studied the mutual interaction of a ferromagnetic
spin-1 condensate with a single-mode cavity. The intracavity light field and
condensate wavefunctions are calculated self-consistently. The
cavity-mediated effective interaction gives rise to a variety of new
spin-depedent phenomena. Under the SMA, both the equilibrium properties and
non-equilibrium dynamics are investigated in detail. We show that the system
can display bistable behavior. By tuning the cavity-pump detuning, the
spin-mixing dynamics can be manipulated. We also discussed the situation
when the SMA becomes invalid, and found that phase transition among
different spin components can occur in the ground state which leads to spin
domain structure. All these effects can be readily tested in experimens. The
cavity-spinor condensate system can provide a new platform for the study of
cavity nonlinear optics and the properties of spinor condensates.

\begin{acknowledgments}
This work is supported by the National Natural Science Foundation of China
under Grant No. 10588402, the National Basic Research Program of China (973
Program) under Grant No. 2006CB921104, the Program of Shanghai Subject Chief
Scientist under Grant No. 08XD14017, Shanghai Leading Academic Discipline
Project under Grant No. B480 (W.Z.), and by the NSFC, \textquotedblleft Chen
Guang\textquotedblright\ project supported by Shanghai Municipal Education
Commission and Shanghai Education Development Foundation (L.Z.), the
Fundamental Research Funds for the Central Universities (L.Z., K.Z.), and by
the NSF (H.P., H.Y.L.), ARO (H.Y.L.), and the Welch Foundation with grant
C-1669 (H.P.).
\end{acknowledgments}

\appendix

%

\section{derivation of $\mathcal{M}$}

Inserting $\psi _{\alpha }=\left( \sqrt{n_{\alpha }^{s}}+\delta \psi
_{\alpha }\right) \exp \left( -i\mu _{\alpha }^{s}t-i\theta _{\alpha
}^{s}\right) $ and $\alpha =\alpha ^{s}+\delta \alpha $ into Eqs. (\ref{gea}%
) and (\ref{geb}) where $\alpha ^{s}$ is the steady-state value of Eq. (\ref%
{cavity}) corresponding to the equilibrium solutions $\left( n_{\alpha
}^{s},\theta _{\alpha }^{s}\right) $, in the homogeneous case ($V_{T}=0$),
keeping terms up to first order in $\delta \psi _{\alpha }$ and $\delta
\alpha $, we obtain
\begin{widetext}
\begin{align}
i\hbar \delta \dot{\psi}_{\pm }& =\left[ -\hbar ^{2}\nabla ^{2}/2m_{a}-\mu
_{\pm }^{s}+U_{0}\left\vert \alpha ^{s}\right\vert ^{2}+2\left(
c_{0}+c_{2}\right) n_{\pm }^{s}+\left( c_{0}+c_{2}\right) n_{0}^{s}+\left(
c_{0}-c_{2}\right) n_{\mp }^{s}\right] \delta \psi _{\pm }  \notag \\
& +\left[ \left( c_{0}+c_{2}\right) n_{\pm }^{s}+c_{2}n_{0}^{s}\exp \left(
-i\theta ^{s}\right) \right] \delta \psi _{\pm }^{\ast }+\left(
c_{0}+c_{2}\right) \sqrt{n_{0}^{s}n_{\pm }^{s}}\left( \delta \psi
_{0}+\delta \psi _{0}^{\ast }\right)   \notag \\
& +\left( c_{0}-c_{2}\right) \sqrt{n_{-}^{s}n_{+}^{s}}\left( \delta \psi
_{\mp }+\delta \psi _{\mp }^{\ast }\right) +2c_{2}\sqrt{n_{0}^{s}n_{\mp }^{s}%
}\delta \psi _{0}\exp \left( -i\theta ^{s}\right)   \notag \\
& +U_{0}\sqrt{n_{\pm}^{s}}\left( \alpha ^{s}\delta \alpha ^{\ast
}+\alpha ^{s\ast }\delta \alpha \right) ,  \label{eea}
\end{align}%
%
\begin{align}
i\hbar \delta \dot{\psi}_{0}& =\left[ -\hbar ^{2}\nabla
^{2}/2m_{a}-\mu _{0}^{s}+\left( c_{0}+c_{2}\right) \left(
n_{+}^{s}+n_{-}^{s}\right) +2c_{0}n_{0}^{s}\right] \delta \psi
_{0}+c_{0}n_{0}^{s}\delta \psi
_{0}^{\ast }  \notag \\
& +\left( c_{0}+c_{2}\right) \left[ \sqrt{n_{+}^{s}n_{0}^{s}}\left( \delta
\psi _{+}+\delta \psi _{+}^{\ast }\right) +\sqrt{n_{-}^{s}n_{0}^{s}}\left(
\delta \psi _{-}+\delta \psi _{-}^{\ast }\right) \right]   \notag \\
& +2c_{2}\left( \sqrt{n_{+}^{s}n_{-}^{s}}\delta \psi _{0}^{\ast }+\sqrt{%
n_{+}^{s}n_{0}^{s}}\delta \psi _{-}+\sqrt{n_{-}^{s}n_{0}^{s}}\delta \psi
_{+}\right) \exp \left( i\theta ^{s}\right) ,  \label{eeb}
\end{align}%
\end{widetext}
and
\begin{align}
\delta \alpha & =-\frac{iU_{0}V\alpha ^{s}}{\kappa -i\left[ \delta
_{c}-U_{0}\left( N_{+}^{s}+N_{-}^{s}\right) \right] }\left[ \sqrt{n_{+}^{s}}%
\left( \delta \psi _{+}+\delta \psi _{+}^{\ast }\right) \right.  \notag \\
& \left. +\sqrt{n_{-}^{s}}\left( \delta \psi _{-}+\delta \psi _{-}^{\ast
}\right) \right] ,  \label{cavity fluctuation}
\end{align}%
where $V=N/n$ is the volumn of the condensate and the use of Eq. (\ref%
{cavity}) has been made in arriving at Eq. (\ref{cavity fluctuation}).
Finally, by combining Eqs. (\ref{eea}), (\ref{eeb}) and (\ref{cavity
fluctuation}), we can construct matrix $\mathcal{M}$ in a straightforward
way.


\begin{thebibliography}{99}
\bibitem{ketterle98} J. Stenger, S. Inouye, D. M. Stamper-Kurn, H.-J.
Miesner, A. P. Chikkatur, and W. Ketterle, Nature (London) \textbf{396}, 345
(1998).

\bibitem{stamper06} L. E. Sadler, J. M. Higbie, S. R. Leslie, M.
Vengalattore, and D. M. Stamper-Kurn, Nature (London) \textbf{443}, 312
(2006).

\bibitem{chapman05} M.-S. Chang, Q. S. Qin, W. Zhang, L. You, and M. S.
Chapman, Nat. Phys. \textbf{1}, 111 (2005).

\bibitem{chapman04} M.-S. Chang, C. D. Hamley, M. D. Barrett, J. A. Sauer,
K. M. Fortier, W. Zhang, L. You, and M. S. Chapman, Phys. Rev. Lett. \textbf{%
92}, 140403 (2004).

\bibitem{lett07} A. T. Black, E. Gomez, L. D. Turner, S. Jung, and P. D.
Lett, Phys. Rev. Lett. \textbf{99}, 070403 (2007).

\bibitem{ketterle03} A. E. Leanhardt, Y. Shin, D. Kielpinski, D. E.
Pritchard, and W. Ketterle, Phys. Rev. Lett. \textbf{90}, 140403 (2003).

\bibitem{ketterle99} H.-J. Miesner, D. M. Stamper-Kurn, J. Stenger, S.
Inouye, A. P. Chikkatur, and W. Ketterle, Phys. Rev. Lett. \textbf{82}, 2228
(1999).

\bibitem{lett09} Y. Liu, S. Jung, S. E. Maxwell, L. D. Turner, E. Tiesinga,
and P. D. Lett, Phys. Rev. Lett. \textbf{102}, 125301 (2009).

\bibitem{ho98} T.-L. Ho, Phys. Rev. Lett. \textbf{81}, 742 (1998).

\bibitem{damski07} B. Damski and W. H. Zurek, Phys. Rev. Lett. \textbf{99},
130402 (2007); New J. Phys. \textbf{11}, 063014 (2009).

\bibitem{pu00} H. Pu, S. Raghavan, and N. P. Bigelow, Phys. Rev. A \textbf{61%
}, 023602 (2000).

\bibitem{you05} W. Zhang, D. L. Zhou, M.-S. Chang, M. S. Chapman, and L.
You, Phys. Rev. A \textbf{72}, 013602 (2005).

\bibitem{passos04} D. R. Romano and E. J. V. de Passos, Phys. Rev. A \textbf{%
70}, 043614 (2004).

\bibitem{jordi09} J. Mur-Petit, Phys. Rev. A \textbf{79}, 063603 (2009).

\bibitem{pu982} C. K. Law, H. Pu, and N. P. Bigelow, Phys. Rev. Lett.
\textbf{81}, 5257 (1998).

\bibitem{santos07} L. Santos, M. Fattori, J. Stuhler, and T. Pfau, Phys.
Rev. A \textbf{75}, 053606 (2007).

\bibitem{kivshar01} N. P. Robins, W. Zhang, E. A. Ostrovskaya, and Y. S.
Kivshar, Phys. Rev. A \textbf{64}, 021601(R) (2001).

\bibitem{ueda05} H. Saito and M. Ueda, Phys. Rev. A \textbf{72}, 023610
(2005).

\bibitem{zhang05} W. Zhang, D. L. Zhou, M.-S. Chang, M. S. Chapman, and L.
You, Phys. Rev. Lett. \textbf{95}, 180403 (2005).

\bibitem{kivshar08} M. Matuszewski, T. J. Alexander, and Y. S. Kivshar,
Phys. Rev. A \textbf{78}, 023632 (2008); Phys. Rev. A \textbf{80}, 023602
(2009).

\bibitem{gu07} Q. Gu and H. Qiu, Phys. Rev. Lett. \textbf{98}, 200401
(2007); C. Tao and Q. Gu, Phys. Rev. A \textbf{79}, 023612 (2009).

\bibitem{esslinger07} F. Brennecke, T. Donner, S. Ritter, T. Bourdel, M. K%
\"{o}hl, and T. Esslinger, Nature \textbf{450}, 268 (2007).

\bibitem{reichel07} Y. Colombe, T. Steinmetz, G. Dubois, F. Linke, D.
Hunger, and J. Reichel, Nature \textbf{450}, 272 (2007).

\bibitem{kurn07} S. Gupta, K. L. Moore, K. W. Murch, and D. M. Stamper-Kurn,
Phys. Rev. Lett. \textbf{99}, 213601 (2007).

\bibitem{kurn08} K. W. Murch, K. L. Moore, S. Gupta, and D. M. Stamper-Kurn,
Nature Phys. \textbf{4}, 561 (2008).

\bibitem{esslinger08} F. Brennecke, S. Ritter, T. Donner, and T. Esslinger,
Science \textbf{322}, 235 (2008); S. Ritter, F. Brennecke, K. Baumann, T.
Donner, C. Guerlin, and T. Esslinger, Appl. Phys. B \textbf{95}, 213 (2009).

\bibitem{ritsch00} P. Horak, S. M. Barnett, and H. Ritsch, Phys. Rev. A
\textbf{61}, 033609 (2000); P. Horak and H. Ritsch, Phys. Rev. A \textbf{63}%
, 023603 (2001).

\bibitem{domokos02} P. Domokos and H. Ritsch, Phys. Rev. Lett. \textbf{89},
253003 (2002); G. Szirmai and P. Domokos, Euro. Phys. J. D \textbf{48}, 127
(2008).


\bibitem{zhang08} J. M. Zhang, W. M. Liu, and D. L. Zhou, Phys. Rev. A
\textbf{77}, 033620 (2008); Phys. Rev. A \textbf{78}, 043618 (2008).

\bibitem{moore99} M. G. Moore and P. Meystre, Phys. Rev. A \textbf{59},
R1754 (1999); M. G. Moore, O. Zobay, and P. Meystre, Phys. Rev. A \textbf{60}%
, 1491 (1999).

\bibitem{larson08} J. Larson, B. Damski, G. Morigi, and M. Lewenstein, Phys.
Rev. Lett. \textbf{100}, 050401 (2008); J. Larson S. Fern\'{a}ndez-Vidal, G.
Morigi, and M. Lewenstein, New J. Phys. \textbf{10}, 045002 (2008).

\bibitem{meystre10} R. Kanamoto and P. Meystre, Phys. Rev. Lett. \textbf{104}%
, 063601 (2010); K. Zhang, W. Chen, M. Bhattacharya, and P. Meystre, Phys.
Rev. A \textbf{81}, 013802 (2010); W. Chen, K. Zhang, D. S. Goldbaum, M.
Bhattacharya, and P. Meystre, Phys. Rev. A \textbf{80}, 011801(R) (2009).

\bibitem{mekhov07} I. B. Mekhov, C. Maschler, and H. Ritsch, Nat. Phys.
\textbf{3}, 319 (2007); I. B. Mekhov, C. Maschler, H. Ritsch, Phys. Rev.
Lett. \textbf{98}, 100402 (2007); I. B. Mekhov and H. Ritsch, Phys. Rev.
Lett. \textbf{102}, 020403 (2009).

\bibitem{zhou09} L. Zhou, H. Pu, H. Y. Ling, and W. Zhang, Phys. Rev. Lett.
\textbf{103}, 160403 (2009).

\bibitem{zhou07} L. Zhou, W. Zhang, H. Y. Ling, L. Jiang, and H. Pu, Phys.
Rev. A \textbf{75}, 043603 (2007).

\bibitem{boyd} R. W. Boyd, \textit{Nonlinear Optics} (Academic Press,
Boston, 2003), 2nd ed.

\bibitem{lambda} $\lambda _{a}$ is estimated with $a_{0}=101.8a_{B}$, $%
a_{2}=100.4a_{B}$ for $^{87}$Rb atoms, the cavity decay rate $\kappa =\left(
2\pi \right) 100$ KHz\ and a typical atomic density $n=1.9\times 10^{14}$ cm$%
^{-3}$.

\bibitem{meystre79} R. Bonifacio and P. Meystre, Opt. Comm. \textbf{29}, 131
(1979); K. Zhang, W. Chen, and P. Meystre, Opt. Comm. \textbf{283}, 665
(2010).
\end{thebibliography}
\end{document}